# Information theoretic security by the laws of classical physics

(INVITED PAPER)

**R. Mingesz [1], L.B. Kish [2], Z. Gingl [1], C.G. Granqvist [3], H. Wen [2,4],**

**F. Peper [5], T. Eubanks [6], G. Schmera [7]**

[1] Department of Technical Informatics, University of Szeged, Árpád tér 2, Szeged, H-6701, Hungary

[2] Texas A&M University, Department of Electrical and Computer Engineering, College Station, TX 77843-3128, USA

[3] Department of Engineering Sciences, The Ångström Laboratory, Uppsala University, P.O. Box 534, SE-75121 Uppsala, Sweden

[4] Hunan University, College of Electrical and Information Engineering, Changsha 410082, China

[5] National Institute of Information and Communication Technology, Kobe, Hyogo 651-2492, Japan

[6] Sandia National Laboratories, P.O. Box 5800, Albuquerque, NM 87185-1033, USA

[7] Space and Naval Warfare Systems Center, San Diego, CA 92152, USA

**Abstract** It has been shown recently that the use of two pairs of resistors with enhanced Johnson-noise and a Kirchhoff-loop—*i.e.*, a Kirchhoff-Law-Johnson-Noise (KLJN) protocol—for secure key distribution leads to information theoretic security levels superior to those of a quantum key distribution, including a natural immunity against a man-in-the-middle attack. This issue is becoming particularly timely because of the recent full cracks of practical quantum communicators, as shown in numerous peer-reviewed publications. This presentation first briefly surveys the KLJN system and then discusses related, essential questions such as: what are perfect and imperfect security characteristics of key distribution, and how can these two types of securities be unconditional (or information theoretical)? Finally the presentation contains a live demonstration.

## 1. Introduction: quantum security hacked

Practical quantum communicators—including several commercial ones—have been fully cracked, as shown in numerous recent papers [1-15], and Vadim Makarov, who is one of the leading quantum crypto crackers, says in *Nature News* that "Our hack gave 100% knowledge of the key, with zero disturbance to the system" [1]. This claim hits at the foundations of quantum encryption schemes because the basis of the security of quantum key distribution (QKD) protocols is the assumption that any eavesdropper (Eve) will disturb the system enough to be detected by the communicator parties (Alice and Bob). Furthermore this proves that we were right in 2007 when claiming in our *SPIE Newsroom* article [16] that quantum security is mainly theoretical because, at that time, no effort had been made to experimentally crack the communicators; instead research grants supported the development of new QKD schemes but not the "politically incorrect" challenge to crack them.

However, the last few years have seen a radical changed the picture [1-15] on the security of practical quantum communicators, and even a full-field implementation of a perfect eavesdropper on a quantum cryptography system has been carried out [2], which is a most difficult task and is an attack on an already established "secure" QKD connection. These cracking schemes are referred to as "hacking" because they utilize physical non-idealities in the building elements of QKD devices. The number of these non-idealities is large, and so is the number of hacking types. The key lessons that has been learned here are that

(*i*) Quantum security at the moment is theoretical, and the applied theory is incorrect for practical devices; a new defense mechanism must be developed for each type of hacking attack, and the potential for yet unexplored non-idealities/ attacks is huge, and

(*ii*) Security analysis, taking into the account of the real physics of the devices, is essential when security matters.

An important aspects all these quantum attacks is the extraordinary (100%) success ratio (*i.e.*, information leak) of extracting the "secure" key bits by Eve, while Alice and Bob do not have a clue that efficient eavesdropping is going on. At this point we note that this information leak was only 0.19% for the *classical* secure communication scheme we are discussing in this paper in the case of a similar situation wherein the strongest vulnerability based on physical non-idealities was used; this is discussed further below.

Inspired by these interesting developments we discuss related issues in the key exchange system of the classical physical Kirchhoff-Law-Johnson-Noise (KLJN) protocol [16]. It should be noted here that there is a general misunderstanding of the KLJN scheme among people lacking the relevant expertise in statistical physics and noise-in-circuitry, as evidenced for example in the Wikipedia entry "Kish cypher" and its "talk page" where, most of the time, both the supporters and the opponents are wrong and the debate falls very short of an objective scientific discussion (amusingly, even the name "cypher" is incorrect). Therefore, after briefly surveying the KLJN system and its properties, we clarify the meaning of *perfect security* and *imperfect security* levels and also define the conditions of these measures: *information theoretic security* (or *unconditional security*) and its limited version *computationally unconditional security*. Furthermore we mention existing integer-number-based key exchange protocols that have (computationally) *conditional security*. It will be seen that theoretical/ideal QKD and KLJN protocols have perfect information theoretic (unconditional) security. However these schemes, when realized with practical/realistic (physical/non-ideal) building elements have imperfect security that is still information theoretic (unconditional), even though current QKD cracks [1-15] indicate that KLJN performs better.

## 2. The KLJN secure key exchange protocol

It is often believed that quantum physics represents modern science and that classical physics is old and outdated. Of course this is not true because the two fields rather pertain to different physical size regimes—the "small" versus the "large" where the appropriate rules of physics are different—not different periods of science history. The above claim regarding "modern" and "old" cannot be maintained even for the history of physics, though, when the point at issue concerns spontaneous random fluctuation phenomena, that are simply referred to as "noise", and it is true for even the most general and omnipresent type of classical physical noise, *viz*., thermal noise (voltage or current fluctuations in thermal equilibrium) which is a younger field of physics than quantum mechanics. Indeed two Swedish scientists, John Johnson and Harry Nyquist both working at Bell Labs, discovered/explained the thermal noise voltage of resistors [17,18] several years after the completion of the foundations of quantum physics [19].

Similarly, quantum heat engines [20] with optional internal coherence effects [21] were proposed several years earlier than the application [22] of the thermal noise of resistors for a heat engine scheme with similar coherence effects.

Finally, the application of thermal noise for unconventional informatics, namely for noise-based logic and computing [23-30] and the KLJN secure key exchange [31-46], emerged decades later than the corresponding quantum informatics schemes such as quantum computing [47] and quantum encryption [48-50].

It is interesting to not that some "exotic" phenomena previously thought to belong to the class of "quantum-weirdness" occur and can be utilized also in the noise schemes, for example: teleportation/telecloning in KLJN networks [45] and entanglement in noise-based logic [23-30].

### *2.1 The Kirchhoff-Law-Johnson-Noise key distribution* [1]

---

[1] This section is a modified version of related expositions elsewhere [36,46].

The KLJN secure key exchange scheme was introduced in 2005 [31-33] and was built and demonstrated in 2007 [34]; it is founded on the robustness of classical information as well as stochasticity and the laws of classical physics. It was named by its creators the "Kirchhoff-loop-Johnson(-like)-Noise" scheme, while on the internet—in blogs and similar sites, including Wikipedia—it has widely been nicknamed "Kish cypher" or "Kish cipher" (where both designations are wrong). The concept has often been misinterpreted and misjudged.

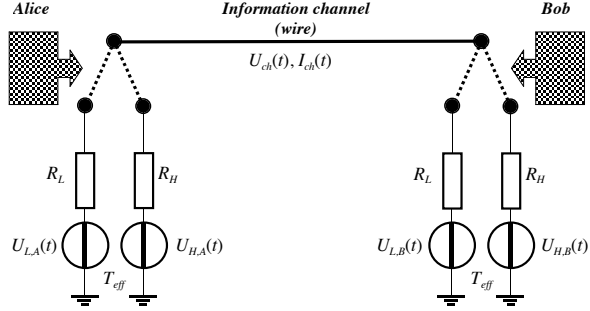

**Figure 1.** Core of the KJLN secure key exchange system [31]. In the text below, the mathematical treatment is based on the power density spectra of the voltages and currents shown in the figure.

The KLJN scheme is a statistical-physical competitor to quantum communicators whose security is based on Kirchhoff's Loop Law and the Fluctuation-Dissipation Theorem. More generally, it is founded on the Second Law of Thermodynamics, which indicates that the security of the ideal scheme is as strong as the impossibility to build a perpetual motion machine of the second kind.

We first briefly survey the foundations of the KLJN system [31,33,36]. Figure 1 shows a model of the idealized KLJN scheme designed for secure key exchange [31]. The resistors $R_L$ and $R_H$ represent the low, $L$ (0), and high, $H$ (1), bits, respectively. At each clock period, Alice and Bob randomly choose one of the resistors and connect it to the wire line. The situation $LH$ or $HL$ represents secure bit exchange [31], because Eve cannot distinguish between them through measurements, while $LL$ and $HH$ are insecure. The Gaussian voltage noise generators (white noise with publicly agreed bandwidth) represent a corresponding thermal noise at a publicly agreed effective temperature $T_{eff}$ (typically $T_{eff} > 10^9$ K [34]). According to the Fluctuation-Dissipation Theorem, the power density spectra $S_{u,L}(f)$ and $S_{u,H}(f)$ of the voltages $U_{L,A}(t)$ and $U_{L,B}(t)$ supplied by the voltage generators in $R_L$ and $R_H$ are given by

$$S_{u,L}(f) = 4kT_{eff}R_L \quad \text{and} \quad S_{u,H}(f) = 4kT_{eff}R_H \quad , \tag{1}$$

respectively.

In the case of secure bit exchange (*i.e.*, the *LH* or *HL* situation), the power density spectrum of channel voltage $U_{ch}(t)$ and channel current $I_{ch}(t)$ are given as

$$S_{u,ch}(f) = 4kT_{eff} \frac{R_L R_H}{R_L + R_H} \quad , \tag{2}$$

and

$$S_{i,ch}(t) = \frac{4kT_{eff}}{R_L + R_H} \quad ; \tag{3}$$

further details are given elsewhere [31,36]. It should be observed that during the *LH* or *HL* case, linear superposition turns Equation (2) into the sum of the spectra of two situations, *i.e.*, when only the generator in $R_L$ is running one gets

$$S_{L,u,ch}(f) = 4kT_{eff} R_L \left(\frac{R_H}{R_L + R_H}\right)^2 \quad , \tag{4}$$

and when the generator in $R_H$ is running one has

$$S_{H,u,ch}(f) = 4kT_{eff} R_H \left(\frac{R_L}{R_L + R_H}\right)^2 \quad . \tag{5}$$

The ultimate security of the system against passive attacks is provided by the fact that the power $P_{H \to L}$, by which the Johnson noise generator of resistor $R_H$ is heating resistor $R_L$, is equal to the power $P_{L \to H}$ by which the Johnson noise generator of resistor $R_L$ is heating resistor $R_H$ [31,36]. A proof of this can also be derived from Equation (3) for a frequency bandwidth of $\Delta f$ by

$$P_{L \to H} = \frac{S_{L,u,ch}(f)\Delta f}{R_H} = 4kT_{eff} \frac{R_L R_H}{(R_L + R_H)^2} \quad , \tag{6a}$$

and

$$P_{H \to L} = \frac{S_{H,u,ch}(f)\Delta f}{R_L} = 4kT_{eff} \frac{R_L R_H}{(R_L + R_H)^2} \quad . \tag{6b}$$

The equality $P_{H \to L} = P_{L \to H}$ (*cf.* Equations 6) is in accordance with the Second Law of Thermodynamics; violating this equality would mean not only going against basic laws of physics and the inability to build a perpetual motion machine (of the second kind) but also allow Eve to use the voltage-current cross-correlation $\langle U_{ch}(t)I_{ch}(t)\rangle$ to extract the bit [31]. However $\langle U_{ch}(t)I_{ch}(t)\rangle = 0$, and hence Eve has an insufficient number of independent equations to determine the bit location during the *LH* or *HL* situation. The above security proof against passive (listening) attacks holds only for Gaussian noise, which has the well-known property that its power density spectrum or autocorrelation function provides the maximum information about the noise and no higher order distribution functions or other tools are able to contribute additional information.

It should be observed [31,33,34,36] that deviations from the shown circuitry—including parasitic elements, inaccuracies, non-Gaussianity of the noise, *etc*.—will cause a potential information leak toward Eve. One should note that the circuit symbol "line" in the circuitry represents an ideal wire with uniform instantaneous voltage and current along it. Thus if the wire is so long and the frequencies are so high that waves appear in it, this situation naturally means that the actual circuitry deviates from the ideal one because neither the voltage nor the current is uniform along the line [31].

To provide unconditional security against invasive attacks, including the man-in-the-middle attack, the fully armed KLJN system shown in Figure 2 monitors the instantaneous current and voltage values at both ends (*i.e*., for Alice as well as Bob) [33,34,36], and these values are compared either via broadcasting them or via an authenticated public channel. An alarm goes off whenever the circuitry is changed or tampered with or energy is injected into the channel. It is important to note that these current and voltage data contain all of the information Eve can possess. This implies that Alice and Bob have full knowledge about the information Eve may have; this is a particularly important property of the KLJN system, which can be utilized in secure key exchange.

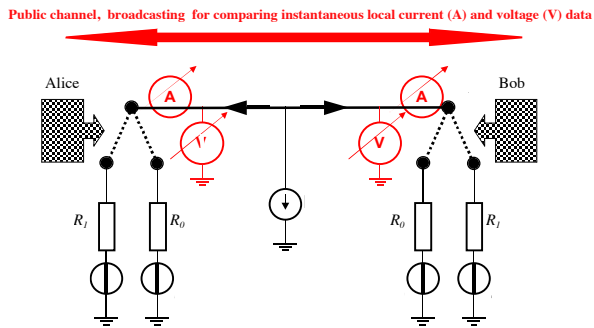

**Figure 2.** Sketch of the KLJN wire communication arrangement [33,36]. To detect the invasive eavesdropper (represented, for example, by the current generator at the middle), the instantaneous current and voltage data measured at the two ends are broadcasted and compared. The eavesdropping is detected immediately, within a small fraction of the time needed to transfer a single bit. Thus statistics of bit errors is not needed, so the exchange of even a single key bit is secure.

The situation discussed above implies the following important features of the KLJN system [31,33,34,36]:

(1) In a practical (non-idealized) KLJN system, Eve can utilize device non-idealities to extract some of the information by proper measurements. This is measurement information and does not depend on Eve's computational and algorithmic ability, *i.e.*, the level of security is computationally unconditional. The maximum leak toward Eve can be designed by Alice and Bob by supposing the physically allowed best/ultimate measurement system for Eve. This designed level of security is unconditional in every sense.

(2) Even when the communication is disturbed by invasive attacks or inherent non-idealities in the KLJN arrangement, the system remains secure because no information can be eavesdropped by Eve without the full knowledge of Alice and Bob about this potential incidence, and without the knowledge of the full information that Eve might have extracted (a full analysis of this aspect is provided elsewhere [36]).

(3) In other words, the KLJN system is always secure, even when it is built with non-ideal elements or designed for a non-zero information leak, in the following sense: The current and voltage data inform Alice and Bob about the exact information leak and hence, for each compromised key bit, they can decide to discard it or even to use it to mislead/manipulate Eve [36].

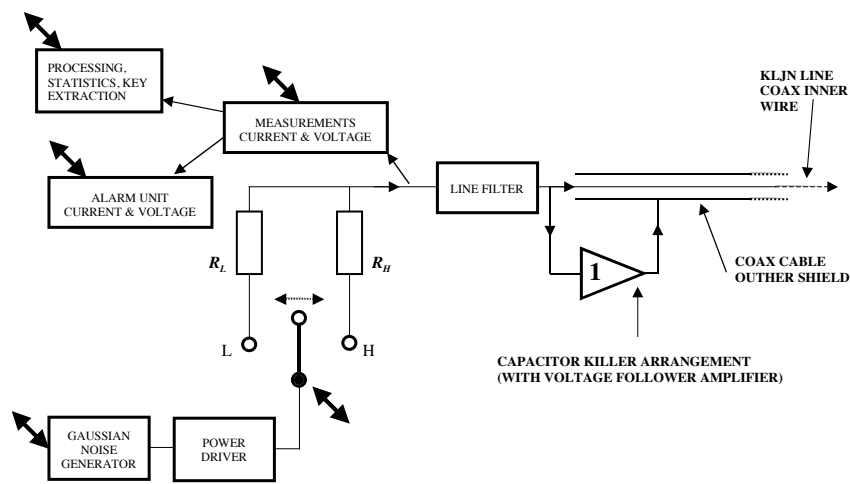

**Figure 3.** A practical KLJN device set-up [34]. Double-ended arrows symbolize computer control.

(4) The KLJN arrangement is naturally and fully protected against the man-in-the-middle attack [33] even during the very first run of the operation when no hidden signatures can be applied. This feature is provided by the unique property of the KLJN system that zero bit information can only be extracted during a man-in-the-middle attack because the alarm goes off before the exchange of a single key bit has taken place [33].

(5) The security of the KLJN system is not based on the error statistics of key bits, and even the exchange of single key bits is secure.

Figure 3 outlines a prototype of the KLJN device [34]. The various non-idealities have been addressed by different tools with the aim that the information leak toward Eve due to non-idealities should stay below 1% of the exchanged raw key bits. For the KLJN device it was 0.19% for the most efficient attack [18]. Here we briefly address two aspects of non-idealities:

(*i*) The role of the line filter (and of the band limitation of the noise generator) is to provide the no-wave limit in the cable, *i.e.*, to preserve the core circuitry (*cf.* Figure 1) in the whole frequency band. This implies that the shortest wavelength component in the driving noise should be much longer than twice the cable length in order to guarantee that no active wave modes and related effects (*e.g.*, reflection, invasive attacks at high frequencies, *etc*.) take place in the cable.

(*ii*) Another tool to fight non-idealities is the cable capacitance compensation ("capacitor killer") arrangement (*cf.* Figure 3). With practical cable parameters and their limits, there is a more serious threat of the security: the cable capacitance shortcuts part of the noise current which results in a greater current at the side of the lower resistance end thus yields an information leak. This effect can be avoided by a cable-capacitor-killer [34] using the inner wire of a coax cable as KLJN line while the outer shield of the cable is driven by the same voltage as the inner wire. However, this is done via a follower voltage amplifier with zero output impedance. The outer shield will then provide all the capacitive currents toward the ground, and the inner wire will experience zero parasitic capacitance. Without "capacitor killer" arrangement and practical bare-wire line parameters, the recommended upper limit of cable length is much shorter and depends on the driving resistor values $R_L$ and $R_H$.

## *2.2. Security proofs and attacks*

The ideal system is absolutely secure, but real systems are rarely ideal and thus hacking attacks are possible by using non-idealities. Fortunately the KLJN system is very simple, implying that the number of such attacks is limited. Several hacking attack types based on the non-ideality of circuit elements causing deviations from the ideal circuitry have been published [36-42]. Each of these attacks triggered a relevant security proof that showed the efficiency of the defense mechanism (*cf.* Figure 2). Furthermore, all known attack types were experimentally tested [34], and the theoretical security proofs were experimentally confirmed.

For practical conditions, the most effective attack employed voltage-drop-related effects on non-zero wire resistance [32,37,38]. It should be noted that serious calculation errors were made by Scheuer and Yariv [37] resulting in a thousand times stronger predicted value of the effect than its real magnitude. The errors were pointed out and the calculations were corrected by Kish and Scheuer [38]. In an experimental demonstration [34], the strongest leak was indeed due to wire resistance, and 0.19% of the bits leaked out ($1.9*10^{-3}$ relative information leak) to Eve, while the fidelity of the key exchange was 99.98% (which means 0.02% bit error rate). This is a very good raw bit leak, and it can easily be made infinitesimally small by simple two-step privacy amplification, as further discussed in Section 2.3.

A general response to the mentioned and other types of small-non-ideality attacks was also presented [39], and the related information leak was shown to be miniscule due to the very poor statistics that Eve could obtain.

Other attack types of less practical significance were based on differences in noise temperatures by Hao [40], which were proven theoretically [41] and experimentally [34] insignificant. The very high accuracy of digital simulations and digital-analog converters (at least 12-bit resolution) allows setting the effective temperature so accurately (0.01% or less error) that this type of inaccuracy-based information leak is not observable. In the case of 12-bit resolution, the theoretical value of the relative information leak is $6*10^{-11}$, *i.e.*, to leak out one effective bit would require a 600 Megabit long key. Therefore this effect was not visible in the experiments even though extraordinarily long (74497) key bits were generated/exchanged in each run [34].

The practical inaccuracy of commercial low-cost resistors (1%) at the two ends [34,41] is a much more serious issue; the theoretical value is <$10^{-4}$ relative information leak (about 7 bits leak from the 74497 bit long key) for a resistance inaccuracy of 1% [34]. However, its impact was still not measurable because of the statistical inaccuracies, $\sqrt{74497} \approx 270$ bits, at this key length. These inaccuracies were about forty times greater than the theoretical information leak of 7 [34].

Wire capacitance would be the most serious source of information leak without the cable-capacitance-killer arrangement, but cable inductance effects are negligible [36].

Another attack [42] focusing on delay effects obtained 70% information leak with a wire simulation software by using physically invalid parameters, such as cable diameters being 28,000 greater than the diameter of the known universe at two km cable length (the error in this attack [42] was pointed out in a subsequent paper [36]). Although this attack was flawed, it is remarkable that even this non-existent, high information leak can be removed by a three-step privacy amplification as discussed in Section 2.3.

It is important to note that the level of allowed information leak is the choice of Alice and Bob, and its actual value is determined only by the invested resources and also typically depends on how much speed is given up. For example, the information leak due to the wire resistance scales inversely with the $6^{th}$ power of wire diameter, which means that employing a ten times thicker cable would reduce the relative information leak of 0.19% to $1.9*10^{-9}$.

For Eve the best attack strategy is to observe the public data exchange about the instantaneous current and voltage amplitudes between Alice and Bob. Those data contain the highest amount of eavesdropping information because they are measured in the most ideal way, and Alice and Bob also base their decision about the bit values on those. Enhancing Eve's infrastructure beyond that ability does not improve her situation, and thus the security is information theoretic/ unconditional.

### *2.3. Privacy amplification in non-ideal systems*

Privacy amplification is a classical software-based technique, which was originally developed for QKD to ensure the security of an encryption scheme with partially exposed key bits. Horvath *et al*. [43] realized simple privacy amplification by executing XOR logic operation on the subsequent pairs of the key bits, thereby halving the key length while progressively reducing the information leak. If the reduction of the information leak is not enough, the same procedure can be repeated on the new key. The resulting key length scales with $0.5^N$, where $N$ is the number of these privacy amplification steps. It was found that, in contrast to quantum key distribution schemes, the high fidelity of the raw key generated in the KLJN system allows the users to always extract a secure shorter key. The necessary conditions are sufficiently high fidelity (small bit error rate), which the KLJN provides, and an upper limit less than one on the eavesdropper probability to correctly guess the exchanged key bits, which means the key exchange is not fully cracked (less than 100% relative information leak is present). The number of privacy amplification steps needed to achieve an information leak of less than $10^{-8}$ in the case of the 0.19% raw bit information leak is two, thus resulting in a corresponding slowdown by a factor of four [43]. In the case of the 70% information leak obtained by the flawed simulations in earlier work [42], the necessary number of privacy amplification steps is three thus resulting in a slowdown of a factor of eight [43].

## 3. Security measures and their conditions

In this section we discuss security measures [52,53] and apply them to compare QKD, KLJN and software security schemes.

A *perfect security* level means that the information channel capacity of the eavesdropping-channel from Alice/Bob toward Eve is zero. *Imperfect security* level means that the information channel capacity of the eavesdropping-channel from Alice/Bob toward Eve is non-zero. We call the encryption "cracked" if Eve can extract all of the information communicated between Alice and Bob. Thus an imperfect security level does not necessarily mean that the encryption is cracked. If the bit-error-rate (BER) is negligible then, by using privacy amplification, the effective level of imperfect security can be enhanced so that it can arbitrarily approach the perfect security level.

To characterize the situations of perfect and imperfect security levels, we must address the conditions where these levels hold. Conditions that both QKD and the KLJN protocols represent are called *information theoretic security*, or *unconditional security*. We note, in passing, that these terms are often completely misunderstood by people who write into Wikipedia and to blog sites about the KLJN system, and these mistakes lead to incorrect conclusions and self-contradicting arguments.

The most rigorous security condition is *information theoretic security*, which means that the information content of the data Eve can extract is limited by information theory even if Eve is using the hypothetical most powerful processing of the extracted data. *Unconditional security* is a similar term indicating security when Eve has unlimited resources. It often means a computationally unconditional security measure, which limits the infrastructure to computers and algorithms, so it has limited validity compared to information theoretic security. Computationally unconditional security simply means that the information content of the data that Eve is able to extract is limited even if she has infinite computing power.

For example, today's generally used software algorithms utilizing prime numbers for key generation and distribution have neither information-theoretic nor computationally unconditional security. All of the information about the key exists in the data observed in the line by Eve, in a decodable form, thus it cannot be information theoretically secure. This information can be fully decoded with a sufficiently fast computer or integer-factoring algorithm, or with a normal computer running for long-enough but finite time. The security is (computationally) conditional: it is based on the assumption that Eve does not have an efficient algorithm or a fast-enough computer to decode the key within the practically relevant time frame.

It is important to note that even imperfect security can be information theoretical or (computationally) unconditional [53]. Such a situation occurs with a physically secure key distribution only, such as QKD or KLJN, because the information leak will be determined by measurement information and not by computation or algorithmic decoding.

The way in which ideal/theoretical QKD makes the key exchange secure is based on the no-cloning theorem of quantum physics: photon states cannot be cloned without introducing errors. Because information bits are carried by (theoretically) single photons, Eve must clone the photon if she wants to measure one; otherwise the information is destroyed before reaching the receiving party. Thus Eve must clone the photon, which introduces extra errors into the line. When Alice and Bob recognize the increased bit-error-rate, they conclude that eavesdropping has happened and they discard the bit-package exhibiting the increased error rate.

The ideal QKD protects the system against eavesdropping, but this is strictly true only for an infinitely long key because Alice and Bob must prepare error statistics, and exact statistics requires infinite time. Otherwise, due to statistical

1212

fluctuations in the BER, Alice and Bob can never be absolutely sure that the key was not eavesdropped. To illustrate this problem, we can go to the simplest type of attacks: the intercept-resend attack for the BB84 QKD protocol (see, for example, [51]). The probability $P(N)$ that the eavesdropping will be discovered while Eve extracts $N$ key bits is not 1 but

$$P_h = 1 - \left(\frac{3}{4}\right)^N \quad . \tag{7}$$

**Table 1.** Comparison of relevant security levels for existing key exchange systems. Practical physically secure key distributions can never have perfect security, they can only approach it.

|  | Perfect | Imperfect | Information theoretic or unconditional | Conditional |
|---|---|---|---|---|
| **QKD theoretical** | **Yes** for the whole key **No** for a single bit | **No** for the whole key **Yes** for a single bit | **Yes** | **No** |
| **KLJN theoretical** | **Yes** for both the whole key and a single bit | **No** | **Yes** | **No** |
| **QKD practical** | **No** | **Yes** | **Yes** | **No** |
| **KLJN practical** | **No** | **Yes** | **Yes** | **No** |
| **Software and prime number based** | **Yes** | **No** | **No** | **Yes** |

Equation (7) shows that, even though a reasonably long key will be very secure and that security can further be enhanced by privacy amplification (see above), the security is not perfect although it can arbitrarily approach the perfect security level. However, if we want to extract only a single key bit, the security is extremely poor because Eve has 25% chance to succeed.

The way by which the ideal/theoretical KLJN scheme makes the key exchange secure depends on the type of the attack: whether it is passive (listening) or invasive (introducing energy in the channel and/or modifying the channel circuitry). In the case of passive listening, information theoretic security due to zero information in the extracted data is guaranteed by the Second Law of Thermodynamics, and this is true even for single-bit attacks where QKD fails. In the case of invasive attacks, the defense mechanics is similar to that of QKD; Alice and Bob will observe deviations between instantaneous signals and they detect the presence of eavesdropping virtually immediately so that, again, even a single bit attack has no chance. Table 1 shows the summary/conclusion about the security level of various key exchange protocols.

In conclusion, the ideal KLJN protocol protects a system against invasive eavesdropping and provides zero information to passive eavesdroppers.

## Acknowledgement

LBK is grateful to Vincent Poor for a discussion about unconditional (information theoretic) security of practical secure physical systems with imperfect security. This work was partially supported by grant TAMOP-4.2.1/B-09/1/KONV-2010-0005. HW's work was partially supported by the National Natural Science Foundation of China under grant 61002035.

## References

[1] Merali Z (29 August 2009) Hackers blind quantum cryptographers. Nature News, DOI:10.1038/news.2010.436.
[2] Gerhardt I, Liu Q, Lamas-Linares A, Skaar J, Kurtsiefer C, Makarov V (2011) Full-field implementation of a perfect eavesdropper on a quantum cryptography system. Nature Communications 2; article number 349. DOI: 10.1038/ncomms1348.
[3] Lydersen L, Wiechers C, Wittmann C, Elser D, Skaar J, Makarov V (2010) Hacking commercial quantum cryptography systems by tailored bright illumination. Nature Photonics 4:686-689. DOI: 10.1038/NPHOTON.2010.214.
[4] Gerhardt I, Liu Q, Lamas-Linares A, Skaar J, Scarani V, Makarov V, Kurtsiefer C (2011) Experimentally faking the violation of Bell's inequalities. Phys. Rev. Lett. 107:170404. DOI: 10.1103/PhysRevLett.107.170404.
[5] Makarov V, Skaar J (2008) Faked states attack using detector efficiency mismatch on SARG04, phase-time, DPSK, and Ekert protocols. Quantum Information and Computation 8:622-635.
[6] Wiechers C, Lydersen L, Wittmann C, Elser D, Skaar J, Marquardt C, Makarov V, Leuchs G (2011) After-gate attack on a quantum cryptosystem. New J. Phys. 13:013043. DOI: 10.1088/1367-2630/13/1/013043.
[7] Lydersen L, Wiechers C, Wittmann C, Elser D, Skaar J, Makarov V (2010) Thermal blinding of gated detectors in quantum cryptography. Optics Express 18:27938-27954. DOI: 10.1364/OE.18.027938.
[8] Jain N, Wittmann C, Lydersen L, Wiechers C, Elser D, Marquardt C, Makarov V, Leuchs

G (2011) Device calibration impacts security of quantum key distribution. Phys. Rev. Lett. 107:110501. DOI: 10.1103/PhysRevLett.107.110501.

[9] Lydersen L, Skaar J, Makarov V (2011) Tailored bright illumination attack on distributed-phase-reference protocols. J. Mod. Opt. 58:680-685. DOI: 10.1080/09500340.2011.565889.

[10] Lydersen L, Akhlaghi MK, Majedi AH, Skaar J, Makarov V (2011) Controlling a superconducting nanowire single-photon detector using tailored bright illumination. New J. Phys. 13:113042. DOI: 10.1088/1367-2630/13/11/113042.

[11] Lydersen L, Makarov V, Skaar J (2011) Comment on "Resilience of gated avalanche photodiodes against bright illumination attacks in quantum cryptography". Appl. Phys. Lett. 99:196101. DOI: 10.1063/1.3658806.

[12] Sauge S, Lydersen L, Anisimov A, Skaar J, Makarov V (2011) Controlling an actively-quenched single photon detector with bright light. Opt. Express 19:23590-23600.

[13] Lydersen L, Jain N, Wittmann C, Maroy O, Skaar J, Marquardt C, Makarov V, Leuchs G (2011) Superlinear threshold detectors in quantum cryptography. Phys. Rev. Lett. 84:032320. DOI: 10.1103/PhysRevA.84.032320.

[14] Lydersen L, Wiechers C, Wittmann C, Elser D, Skaar J, Makarov V (2010) Avoiding the blinding attack in QKD; REPLY (COMMENT). Nature Photonics 4:801-801. DOI: 10.1038/nphoton.2010.278.

[15] Makarov V (2009) Controlling passively quenched single photon detectors by bright light. New J. Phys. 11:065003. DOI: 10.1088/1367-2630/11/6/065003.

[16] Kish LB, Mingesz R, Gingl Z (2007) Unconditionally secure communication via wire. SPIE Newsroom. DOI: 10.1117/2.1200709.0863.

[17] Johnson JB (1927) Thermal agitation of electricity in conductors. Nature 119:50-51.

[18] Nyquist H (1928) Thermal agitation of electric charge in conductors. Phys. Rev. 32:110-113.

[19] Born M, Heisenberg W, Jordan P (1926) Quantum mechanics II. Z. Phys. 35:557-615.

[20] Allahverdyan AE, Nieuwenhuizen TM (2000) Extraction of work from a single thermal bath in the quantum regime. Phys. Rev. Lett. 85:1799-1802.

[21] Scully MO, Zubairy MS, Agarwal GS, Walther H (2003) Extracting work from a single heat bath via vanishing quantum coherence. Science 299:862–864.

[22] Kish LB (2011) Thermal noise engines. Chaos Solit. Fract. 44:114–121. http://arxiv.org/abs/1009.5942.

[23] Kish LB (2009) Noise-based logic: Binary, multi-valued, or fuzzy, with optional superposition of logic states. Phys. Lett. A 373:911-918.

[24] Kish LB, Khatri S, Sethuraman S (2009) Noise-based logic hyperspace with the superposition of $2^N$ states in a single wire. Phys. Lett. A 373:1928-1934.

[25] Bezrukov SM, Kish LB (2009) Deterministic multivalued logic scheme for information processing and routing in the brain. Phys. Lett. A 373:2338-2342.

[26] Gingl Z, Khatri S, Kish LB (2010) Towards brain-inspired computing. Fluct. Noise Lett. 9:403-412.

[27] Kish LB, Khatri S, Horvath T (2011) Computation using noise-based logic: Efficient string verification over a slow communication channel. Eur. J. Phys. B 79:85-90. http://arxiv.org/abs/1005.1560.

[28] Peper F, Kish LB (2011) Instantaneous, non-squeezed, noise-based logic. Fluct. Noise Lett. 10:231-237. http://www.worldscinet.com/fnl/10/1002/open-access/S0219477511000521.pdf.

[29] Wen H, Kish LB, Klappenecker A, Peper F (2012) New noise-based logic representations to avoid some problems with time complexity. Fluct. Noise Lett. (June 2012 issue, in press); http://arxiv.org/abs/1111.3859.

[30] Mullins J (2010) Breaking the noise barrier. New Scientist, issue 2780 (29 September 2010); http://www.newscientist.com/article/mg20827801.500-breaking-the-noise-barrier.html?full=true.

[31] Kish LB (2006) Totally secure classical communication utilizing Johnson(-like) noise and Kirchhoff's law. Phys. Lett. A 352:178-182.

[32] Cho A (2005) Simple noise may stymie spies without quantum weirdness. Science 309:2148; http://www.ece.tamu.edu/~noise/news_files/science_secure.pdf.
[33] Kish LB (2006) Protection against the man-in-the-middle-attack for the Kirchhoff-loop-Johnson(-like)-noise cipher and expansion by voltage-based security. Fluct. Noise Lett. 6 pp. L57-L63. http://arxiv.org/abs/physics/0512177
[34] Mingesz R, Gingl Z, Kish LB (2008) Johnson(-like)-noise-Kirchhoff-loop based secure classical communicator characteristics, for ranges of two to two thousand kilometers, via model-line, Phys. Lett. A 372:978-984.
[35] Palmer DJ (2007) Noise encryption keeps spooks out of the loop. New Scientist, issue 2605 p.32; http://www.newscientist.com/article/mg19426055.300-noise-keeps-spooks-out-of-the-loop.html
[36] Kish LB, Horvath T (2009) Notes on recent approaches concerning the Kirchhoff-law-Johnson-noise-based secure key exchange. Phys. Lett. A 373:901-904.
[37] Scheuer J, Yariv A (2006) A classical key-distribution system based on Johnson (like) noise – How secure? Phys. Lett. A 359:737-740.
[38] Kish LB, Scheuer J (2010) Noise in the wire: The real impact of wire resistance for the Johnson(-like) noise based secure communicator. Phys. Lett. A 374:2140-2142.
[39] Kish LB (2006) Response to Scheuer-Yariv: "A classical key-distribution system based on Johnson (like) noise – How secure?". Phys. Lett. A 359:741-744.
[40] Hao F (2006) Kish's key exchange scheme is insecure. IEE Proc. Inform. Sec. 153:141-142.
[41] Kish LB (2006) Response to Feng Hao's paper "Kish's key exchange scheme is insecure". Fluct. Noise Lett. 6:C37–C41.
[42] Liu PL (2009) A new look at the classical key exchange system based on amplified Johnson noise. Phys. Lett. A 373:901–904.
[43] Horvath T, Kish LB, Scheuer J (2011) Effective privacy amplification for secure classical communications. Europhys. Lett. 94:28002. http://arxiv.org/abs/1101.4264.
[44] Kish LB, Saidi O (2008) Unconditionally secure computers, algorithms and hardware. Fluct. Noise Lett. 8:L95-L98.
[45] Kish LB, Mingesz R (2006) Totally secure classical networks with multipoint telecloning (teleportation) of classical bits through loops with Johnson-like noise. Fluct. Noise Lett. 6:C9-C21.
[46] Kish LB, Peper F (2012) Information networks secured by the laws of physics. IEICE Trans. Commun. E95-B:1501-1507.
[47] http://en.wikipedia.org/wiki/Quantum computer.
[48] Wiesner S (1983) Conjugate coding, SIGACT News 15:78-88.
[49] Bennett CH, Brassard G (1983) Quantum cryptography and its application to provably secure key expansion, public-key distribution, and coin-tossing. Proc. IEEE Internat. Symp. Inform. Theor., St-Jovite, Canada, p. 91.
[50] Brassard G, (2005) Brief history of quantum cryptography: A personal perspective. Proc. IEEE Information Theory Workshop on Theory and Practice in Information Theoretic Security, Awaji Island, Japan, pp. 19-23, and references therein. http://arxiv.org/abs/quant-ph/0604072.
[51] Xu F, Qi B, Lo HK (2010) Experimental demonstration of phase-remapping attack in a practical quantum key distribution system. New J. Phys. 12:113026. http://arxiv.org/abs/1005.2376.
[52] Liang Y, Poor HV, Shamai S (2008) Information Theoretic Security. Foundations and Trends in Communications and Information Theory 5:355–580. DOI: 10.1561/0100000036.
[53] Vincent Poor, private communications.